\documentclass[prd,aps,nofootinbib,superscriptaddress]{revtex4}
\usepackage{epsfig}
\usepackage{amsmath, slashed}
\usepackage{xcolor}
\usepackage{appendix}
\usepackage{dsfont}

\usepackage[normalem]{ulem}

\newcommand{\uvec}{\boldsymbol}
\newcommand{\ud}{\mathrm{d}}

\usepackage{bm}
\usepackage{amsfonts}

\allowdisplaybreaks
\begin{document}

\title{Gravitational transverse-momentum distributions}

\author{C\'edric Lorc\'e}
\email[]{cedric.lorce@polytechnique.edu}
\affiliation{CPHT, CNRS, \'Ecole polytechnique, Institut Polytechnique de Paris, 91120 Palaiseau, France}

\author{Qin-Tao Song}
\email[]{songqintao@zzu.edu.cn}
\affiliation{CPHT, CNRS, \'Ecole polytechnique, Institut Polytechnique de Paris, 91120 Palaiseau, France}
\affiliation{School of Physics and Microelectronics, Zhengzhou University, Zhengzhou, Henan 450001, China}

\date{\today}

\begin{abstract}
{We study the energy-momentum tensor of spin-$0$ and spin-$\frac{1}{2}$ hadrons in momentum space. We parametrize this object in terms of so-called gravitational transverse-momentum distributions, and we identify in the quark sector the relations between the latter and the usual transverse-momentum distributions. Focusing on particular components of the energy-momentum tensor, we study momentum densities, flux of inertia and stress distribution in momentum space, revealing part of the wealth of physical information that can be gained from higher-twist transverse-momentum distributions.
  }

\end{abstract}

\maketitle

\section{Introduction}

The QCD energy-momentum tensor (EMT) is a key object for studying and understanding the internal structure of hadrons~\cite{Burkert:2023wzr}. It is indeed directly related to the longstanding questions of the hadron mass~\cite{Ji:1994av,Ji:1995sv,Yang:2018nqn,Hatta:2018sqd,Lorce:2017xzd,Metz:2020vxd,Lorce:2021xku} and spin decompositions~\cite{Jaffe:1989jz,Ji:1996ek,Leader:2013jra,Wakamatsu:2014zza,Lorce:2021gxs}. Moreover, it allows one to investigate the mechanical properties of hadrons~\cite{Polyakov:2002yz,Polyakov:2018zvc,Burkert:2018bqq,Lorce:2018egm,Freese:2021czn}. Studying the EMT is therefore of prime importance and stands at the heart of the physics program of the future Electron-Ion Collider in the US~\cite{AbdulKhalek:2021gbh,AbdulKhalek:2022hcn}.

Direct access to the EMT requires a gravitational probe, and is in practice out of reach owing to the extreme weakness of gravitational interactions at the microscopic level. Fortunately, in QCD the EMT can be probed indirectly via electromagnetic interactions. Matrix elements of the local EMT operator have been parametrized in terms of gravitational form factors~\cite{Kobzarev:1962wt,Pagels:1966zza,Ji:1996ek,Bakker:2004ib,Cotogno:2019vjb}. The latter can then be related to generalized parton distributions (GPDs)~\cite{Ji:1996ek} and generalized distribution amplitudes~\cite{Kumano:2017lhr} accessible in various experimental processes, see e.g.~\cite{Diehl:2003ny}. This has been generalized to the case of a non-local EMT operator, whose general matrix elements have been parametrized in terms of what can be called gravitational GPDs\footnote{Strictly speaking, the scalar functions introduced in Ref.~\cite{Lorce:2015lna} correspond to gravitational GPDs integrated over the parton longitudinal momentum, but the general parametrization is not impacted by this integration since we considered non-local EMT operators.}~\cite{Lorce:2015lna}. Similar objects have later been considered in Ref.~\cite{Guo:2021aik}.

While the connection between the EMT and GPDs is well established, the link with another class of non-perturbative functions known as transverse-momentum distributions (TMDs)~\cite{Boer:1997nt} has so far been limited to the longitudinal and transverse momentum sum rules~\cite{Burkardt:2003yg,Burkardt:2004ur,Lorce:2015lna,Boer:2015vso,Amor-Quiroz:2020qmw}. The aim of the present work is to introduce the notion of EMT distribution in momentum space and to identify the physical information about the EMT that can be accessed via TMDs. The paper is organized as follows. In Section \ref{sect2} we define the transverse-momentum dependent EMT and we parametrize the associated matrix elements in terms of gravitational TMDs. We then discuss in Section~\ref{sect3} the connection with the standard quark TMDs and we study in Section~\ref{sect4} part of the physical content that can be accessed from twist-2 and twist-3 TMDs. Finally, we summarize our findings in Section~\ref{sect5}.

\section{Gravitational TMDs}
\label{sect2}

\subsection{TMD correlator}\label{sec-correlators}

 We start with a reminder on the TMD correlators. The fully unintegrated quark-quark correlator for a spin-$\frac{1}{2}$ target~\cite{Meissner:2009ww} is defined in the forward limit as
\begin{equation}\label{forwardGTMD}
    W^{[\Gamma]}(P,k,N,S;\eta)=\frac{1}{2}\int\frac{\ud^4z}{(2\pi)^4}\,e^{ik\cdot z}\,\langle P,S|\overline\psi(-\tfrac{z}{2})\Gamma\mathcal W(-\tfrac{z}{2},\tfrac{z}{2}|n)\psi(\tfrac{z}{2})|P,S\rangle,
\end{equation}
where $\Gamma$ stands for a generic matrix in Dirac space, e.g. $\Gamma=\gamma^\mu,\gamma^\mu\gamma_5,\cdots$. For a target of mass $M$ and four-momentum $P$, the covariant spin vector $S$ defined via $\overline u(P,S)\gamma^\mu\gamma_5u(P,S)=2MS^\mu$ satisfies $P\cdot S=0$ and $S^2=-1$. The quark average four-momentum $k$ is defined as the Fourier conjugate variable to the space-time distance $z$ between the two quark operators. Gauge invariance is preserved by the inclusion of a Wilson line $\mathcal W$ connecting the points $-\frac{z}{2}$ and $\frac{z}{2}$ via an infinitely long staple-shaped path along the lightlike direction $n$. Since the same Wilson line is unchanged under the rescaling $n\mapsto \alpha n$ with $\alpha>0$, the correlator depends in fact on the rescaling-invariant four-vector
\begin{equation}
    N=\frac{M^2n}{P\cdot n}.
\end{equation}
The parameter $\eta=\text{sign}(n^0)$ indicates whether the Wilson line is future-pointing ($\eta=+1$) or past-pointing ($\eta=-1$).

For convenience, we choose the coordinate system and the rescaling factor $\alpha$ such that
\begin{equation}
\begin{aligned}
    P^\mu&=\left[P^+,\frac{M^2}{2P^+},\uvec 0_\perp\right],\\
    k^\mu&=\left[xP^+,k^-,\uvec k_\perp\right],\\
    n^\mu&=\left[0,\eta,\uvec 0_\perp\right],
\end{aligned}
\end{equation}
where $v^\mu=[v^+,v^-,\uvec v_\perp]$ with the light-front components defined as $v^\pm=(v^0\pm v^3)/\sqrt{2}$. The quark TMD correlator (see e.g.~\cite{Bacchetta:2006tn}) is then obtained by integration over the quark light-front energy
\begin{equation}\label{TMDcorr}
\begin{aligned}
    \Phi^{[\Gamma]}(P,x,\uvec k_\perp,N,S;\eta)&=\int\ud k^-\,W^{[\Gamma]}(P,k,N,S;\eta)\\
    &=\frac{1}{2}\int\frac{\ud z^-\,\ud^2z_\perp}{(2\pi)^3}\,e^{ik\cdot z}\,\langle P,S|\overline\psi(-\tfrac{z}{2})\Gamma\mathcal W(-\tfrac{z}{2},\tfrac{z}{2}|n)\psi(\tfrac{z}{2})|P,S\rangle\Big|_{z^+=0}.
\end{aligned}
\end{equation}

\subsection{Transverse-momentum dependent EMT}

In QCD, the local gauge-invariant EMT operator for quarks is given by
\begin{equation}\label{localEMT}
    T^{\mu\nu}_q(r)=\overline\psi(r)\gamma^\mu \tfrac{i}{2}\overset{\leftrightarrow}{D}\!\!\!\!\!\phantom{D}^\nu\psi(r)
\end{equation}
with $\overset{\leftrightarrow}{D}\!\!\!\!\!\phantom{D}^\nu=\overset{\rightarrow}{\partial}\!\!\!\!\phantom{\partial}^\nu-\overset{\leftarrow}{\partial}\!\!\!\!\phantom{\partial}^\nu-2ig A^\nu(r)$. In order to define the EMT for a quark with average four-momentum $k$, we need to consider a bilocal generalization of this expression. Unfortunately, the covariant derivative does not commute with the Wilson line, making the bilocal generalization of Eq.~\eqref{localEMT} ambiguous~\cite{Lorce:2012ce}. The problem can be traced back to the fact that $[D_\mu,D_\nu]\neq 0$ whereas $[k_\mu,k_\nu]=0$, which implies that $k$ cannot be identified with the quark \emph{kinetic} four-momentum. However, if we work in the gauge where the Wilson line reduces to the identity (namely the light-front gauge with appropriate advanced of retarded boundary conditions depending on the value of $\eta$~\cite{Belitsky:2002sm}), the four-vector $k^\mu$ can be represented by the partial derivatives $i\partial^\mu$, and hence be interpreted as the quark \emph{canonical} four-momentum. Therefore, instead of looking for the bilocal generalization of the kinetic EMT operator~\eqref{localEMT}, we should rather be looking for the bilocal generalization of the light-front gauge-invariant canonical (gic) EMT operator~\cite{Lorce:2012rr,Leader:2013jra,Lorce:2015lna}
\begin{equation}
     T^{\mu\nu}_{q,\text{gic}}(r)=\overline\psi(r)\gamma^\mu \tfrac{i}{2}\overset{\leftrightarrow}{D}\!\!\!\!\!\phantom{D}^\nu_\text{pure}\psi(r),
\end{equation}
where $D^\mu_\text{pure}=\partial^\mu-igA^\mu_\text{pure}$ is known as the pure-gauge covariant derivative~\cite{Chen:2008ag,Wakamatsu:2010cb}, corresponding in the present context to the covariant derivative reducing in the light-front gauge $A^+=0$ (with appropriate boundary conditions) to $\partial^\mu$~\cite{Hatta:2011zs,Hatta:2011ku,Lorce:2012ce}. Note that by definition $A^+_\text{pure}(r)=A^+(r)$, meaning that $T^{\mu+}_q(r)=T^{\mu+}_{q,\text{gic}}(r)$. Therefore, as far as the longitudinal light-front momentum is concerned, there is no difference between the kinetic and the gauge-invariant canonical definitions.

Following the spirit of Refs.~\cite{Ji:2003ak,Belitsky:2003nz,Lorce:2011kd,Lorce:2011ni}, it is natural to define the bilocal gauge-invariant canonical (gic) EMT operator for quarks as~\cite{Lorce:2012ce}
\begin{equation}\label{k-moment}
    T^{\mu\nu}_{q,\text{gic}}(r,k)=k^\nu\int\frac{\ud^4z}{(2\pi)^4}\,e^{ik\cdot z}\,\overline\psi(r-\tfrac{z}{2})\gamma^\mu\mathcal W(r-\tfrac{z}{2},r+\tfrac{z}{2}|n)\psi(r+\tfrac{z}{2}).
\end{equation}
Integrating by parts, we can write
\begin{equation}\label{bilocalEMT}
\begin{aligned}
    T^{\mu\nu}_{q,\text{gic}}(r,k)&=\int\frac{\ud^4z}{(2\pi)^4}\,e^{ik\cdot z}\,i\partial^\nu_z\!\left[\overline\psi(r-\tfrac{z}{2})\gamma^\mu\mathcal W(r-\tfrac{z}{2},r+\tfrac{z}{2}|n)\psi(r+\tfrac{z}{2})\right]\\
    &=\int\frac{\ud^4z}{(2\pi)^4}\,e^{ik\cdot z}\left[\overline\psi(r-\tfrac{z}{2})\gamma^\mu\mathcal W(r-\tfrac{z}{2},r+\tfrac{z}{2}|n)\tfrac{i}{2}\overset{\rightarrow}{D}\!\!\!\!\!\phantom{D}^\nu_\text{pure}(r+\tfrac{z}{2})\psi(r+\tfrac{z}{2})\right.\\
    &\qquad\qquad\qquad\quad\left.-\overline\psi(r-\tfrac{z}{2})\tfrac{i}{2}\overset{\leftarrow}{D}\!\!\!\!\!\phantom{D}^\nu_\text{pure}(r-\tfrac{z}{2})\gamma^\mu\mathcal W(r-\tfrac{z}{2},r+\tfrac{z}{2}|n)\psi(r+\tfrac{z}{2})\right],
\end{aligned}
\end{equation}
where $A^\mu_\text{pure}$ is given by
\begin{equation}
    A^\mu_\text{pure}(r)=\mathcal W(r,0|n)\tfrac{i}{g}\partial^\mu_r\mathcal W(0,r|n).
\end{equation}
Since we obviously have the property
\begin{equation}
    D^\mu_\text{pure}(x)\mathcal W(x,y|n)=\mathcal W(x,y|n)D^\mu_\text{pure}(y)
\end{equation}
reflecting the commutativity of pure-gauge covariant derivatives, the bilocal operator in Eq.~\eqref{bilocalEMT} is unambiguous. Moreover, integrating over the quark four-momentum leads to
\begin{equation}
    \int\ud^4k\,T^{\mu\nu}_{q,\text{gic}}(r,k)=T^{\mu\nu}_{q,\text{gic}}(r)
\end{equation}
as expected.

We can now define in a natural way the fully unintegrated EMT by considering the forward matrix element\footnote{The motivation for the factor $\frac{1}{2}$ is the same as for the correlators in Section~\ref{sec-correlators}: the light-front expectation value of an operator $O$ is $\frac{\langle P,S|O|P,S\rangle}{2P^+}$ and switching from a distribution in $k^+$ to a distribution in $x$ amounts to a multiplication by  the Jacobian $P^+$.} of the operator in Eq.~\eqref{bilocalEMT}
\begin{equation}\label{unintegratedEMT}
\Theta^{\mu\nu}_q(P,k,N,S;\eta)=\frac{1}{2}\,\langle P,S|T^{\mu\nu}_{q,\text{gic}}(0,k)|P,S\rangle,
\end{equation}
and the TMD EMT by further integrating over the quark light-front energy
\begin{equation}
\begin{aligned}
\mathcal T^{\mu\nu}_q(P,x,\uvec k_\perp,N,S;\eta)&=\int\ud k^-\, \Theta^{\mu\nu}_q(P,k,N,S;\eta)\\
&=\frac{1}{2}\int\frac{\ud z^-\,\ud^2z_\perp}{(2\pi)^3}\,e^{ik\cdot z}\,i\partial_z^\nu\langle P,S|\overline\psi(-\tfrac{z}{2})\gamma^\mu\mathcal W(-\tfrac{z}{2},\tfrac{z}{2}|n)\psi(\tfrac{z}{2})|P,S\rangle\Big|_{z^+=0}.
\end{aligned}
\end{equation}
This last object can be interpreted as the 3D distribution of the quark EMT in momentum space.

\subsection{Parametrization in terms of gravitational TMDs}

Parity, hermiticity and time-reversal invariance imply that the fully unintegrated EMT satisfies the relations
\begin{equation}\label{eqn:inv}
\begin{aligned}
\Theta^{\mu \nu}(k, P, N, S;\eta)&=\Theta^{\bar{\mu} \bar{\nu}}(\bar{k}, \bar{P}, \bar{N}, -\bar{S};\eta),\\
\Theta^{\mu \nu}(k, P, N, S;\eta)&=[\Theta^{\mu \nu}(k, P, N, S;\, \eta)]^{\dag}, \\
\Theta^{\mu \nu}(k, P, N, S;\eta)&=[\Theta^{\bar{\mu} \bar{\nu}}(\bar{k}, \bar{P}, \bar{N}, \bar{S};-\eta)]^*,
\end{aligned}
\end{equation}
with the notation $v^{\bar{\mu}}= \bar{v}^{\mu}=(v^0,-\uvec{v})$. Since the parametrization should be the same for both quark and gluon contributions to the EMT, we drop the label $q$ in this subsection.

For convenience, we define the transverse part of a four-vector by $v^\mu_T=g^{\mu\nu}_Tv_\nu$ using the projector onto the subspace orthogonal to $P$ and $N$
\begin{align}
g_{T}^{\mu \nu}=g^{\mu \nu}-
\frac{P^{\mu}N^{\nu}+P^{\nu}N^{\mu}}{M^2}+
\frac{N^{\mu}N^{\nu}}{M^2}.
\label{eqn:ktmo1}
\end{align}
The covariant spin vector can then be expressed as
\begin{align}
S^{\mu}=\frac{\lambda}{M} (P^{\mu}-N^{\mu})+S_T^{\mu},
\label{eqn:svector}
\end{align}
where the longitudinal light-front polarization is denoted by the parameter $\lambda$ and the transverse light-front polarization by the four-vector $S^\mu_T=[0,0,\uvec S_\perp]$. We define also the transverse Levi-Civita pseudotensor
\begin{align}
\epsilon_T^{\mu \nu}=\frac{\epsilon^{\mu \nu \alpha \beta} N_{\alpha}P_{\beta}}{M^2}
\label{eqn:aysten}
\end{align}
with the convention $\epsilon_{0123}=1$ such that $\epsilon_T^{12}=1$, and we introduce the compact notation $\epsilon_T^{\mu v_T} \equiv  \epsilon_T^{\mu \nu}  v_{T \nu}$.

A complete parametrization of the TMD EMT $\mathcal{T}^{\mu \nu}(P,x,\uvec k_\perp, N, S; \eta)$ for spin-$0$ and spin-$\frac{1}{2}$ targets can be obtained by writing down all the independent rank-2 tensors built out of $g_T^{\mu\nu}$, $\epsilon_T^{\mu\nu}$, $P^\mu$, $N^\mu$, and $k^\mu_T$, which are at most linear in the polarization and which satisfy the constraints in Eq.~\eqref{eqn:inv}. We find\footnote{Note that other possible tensor structures have been discarded thanks to the Schouten identity
\begin{equation*}
g^{\alpha \beta} \epsilon^{\mu \nu \rho \sigma} +  g^{\alpha \mu} \epsilon^{ \nu \rho \sigma \beta}
+g^{\alpha \nu} \epsilon^{  \rho \sigma \beta \mu} +g^{\alpha \rho} \epsilon^{  \sigma  \beta \mu \nu}
+g^{\alpha \sigma} \epsilon^{   \beta \mu \nu \rho}=0.
\end{equation*} }
\begin{equation}
\begin{aligned}
\mathcal{T}^{\mu \nu}= \frac{1}{P^+} \Big\{& P^{\mu}P^{\nu}a_1+N^{\mu}N^{\nu}a_2+k_T^{\mu}k_T^{\nu}a_3
+P^{\mu}N^{\nu}a_4+N^{\mu}P^{\nu}a_5\\
&+P^{\mu}k_T^{\nu}a_6+k_T^{\mu}P^{\nu}a_7+
N^{\mu}k_T^{\nu}a_8+k_T^{\mu}N^{\nu}a_9+M^2g^{\mu \nu}_Ta_{0}\\
&-\frac{\epsilon_{T}^{k_T  S_T } }{M} \left[ P^{\mu}P^{\nu}a_{1T}^\perp+N^{\mu}N^{\nu}a_{2T}^\perp+k_T^{\mu}k_T^{\nu}a_{3T}^\perp
+P^{\mu}N^{\nu}a_{4T}^\perp+N^{\mu}P^{\nu}a_{5T}^\perp\right.\\
&\qquad\qquad\quad\left.+P^{\mu}k_T^{\nu}a_{6T}^\perp+k_T^{\mu}P^{\nu}a_{7T}^\perp+N^{\mu}k_T^{\nu}a_{8T}^\perp  +k_T^{\mu}N^{\nu}a_{9T}^\perp+ M^2 g^{\mu \nu}_T a_{0T}^\perp \right] \\
&-M \left[
P^{\mu} \epsilon_{T}^{\nu  S_T } a_{1T}+
P^{\nu} \epsilon_{T}^{\mu  S_T } a_{2T}+
N^{\mu} \epsilon_{T}^{\nu  S_T } a_{3T}+
N^{\nu} \epsilon_{T}^{\mu  S_T } a_{4T}+
k_T^{\mu} \epsilon_{T}^{\nu  S_T } a_{5T}+k_T^{\nu} \epsilon_{T}^{\mu  S_T } a_{6T}
\right] \\
&-\lambda  \left[P^{\mu} \epsilon_{T}^{\nu  k_T } a_{1L}+
P^{\nu} \epsilon_{T}^{\mu  k_T } a_{2L}+
N^{\mu} \epsilon_{T}^{\nu  k_T } a_{3L}+
N^{\nu} \epsilon_{T}^{\mu  k_T } a_{4L}+
k_T^{\mu} \epsilon_{T}^{\nu  k_T } a_{5L}+
k_T^{\nu} \epsilon_{T}^{\mu  k_T } a_{6L}
\right] \Big\} ,
\label{eqn:emt1}
\end{aligned}
\end{equation}
where the real-valued coefficients $a_i(x,\uvec k_\perp^2)$ will be referred to as gravitational TMDs. There are 10 polarization-independent gravitational TMDs (viz. $a_{0-9}$). For a spin-$0$ target, that is all we have. For a spin-$\frac{1}{2}$ target, there are in addition 22 polarization-dependent gravitational TMDs: 6 associated with the longitudinal polarization (viz. $a_{1-6L}$) and 16 associated with the transverse polarization (viz. $a_{1-6T}$ and $a_{0-9T}^\perp$). As a result of the discrete symmetries~\eqref{eqn:inv}, the polarization-independent gravitational TMDs are naive \textsf{T}-even (i.e. independent of $\eta$) whereas the polarization-dependent ones are naive \textsf{T}-odd (i.e. they change sign under $\eta\mapsto-\eta$). Interestingly, the same total numbers of gravitational GPDs for spin-$0$ and spin-$\frac{1}{2}$ targets have been obtained in Ref.~\cite{Lorce:2015lna}. Since $\int\ud^2k_\perp\,\mathcal T^{\mu\nu}$ can not depend on $k^\mu_T$, one may naively think by eliminating all the $k^\mu_T$-dependent tensors in Eq.~\eqref{eqn:emt1} that there are only 9 gravitational PDFs. Note however that the combination $k_T^{\mu} \epsilon_{T}^{\nu  k_T }-k_T^{\nu} \epsilon_{T}^{\mu  k_T }=\uvec k_\perp^2\epsilon_T^{\mu\nu}$ does survive integration over $\uvec k_\perp$, meaning that there are in total 10 gravitational PDFs, in agreement with the results in Section 4.4 of Ref.~\cite{Lorce:2015lna}.

\section{Relations between TMDs and gravitational TMDs}
\label{sect3}

In practice, gravitational TMDs cannot be accessed directly in experiments for the scattering amplitudes between hadrons and gravitons are extremely small. Part of them can however be obtained indirectly through their relations with ordinary TMDs. It is easy to see from Eqs.~\eqref{forwardGTMD}, \eqref{k-moment} and \eqref{unintegratedEMT} that at the level of the fully unintegrated matrix elements we have the simple relation
\begin{equation}
    \Theta^{\mu\nu}_q(P,k,N,S;\eta)=k^\nu W^{[\gamma^\mu]}(P,k,N,S;\eta).
\end{equation}
Integrating over $k^-$ leads us to
\begin{equation}\label{reltoTMD}
    \mathcal T^{\mu\nu}_q(P,x,\uvec k_\perp,N,S;\eta)=k^\nu \Phi^{[\gamma^\mu]}(P,x,\uvec k_\perp,N,S;\eta)\qquad\text{for }\nu\neq -.
\end{equation}
Let us therefore consider the quark vector TMD correlator, obtained from Eq.~\eqref{TMDcorr} using $\Gamma=\gamma^\mu$,
\begin{equation}\label{eqn:tmd}
    \Phi^{[\gamma^\mu]}(P,x,\uvec k_\perp,N,S;\eta)=\frac{1}{2}\int\frac{\ud z^-\,\ud^2z_\perp}{(2\pi)^3}\,e^{ik\cdot z}\,\langle P,S|\overline\psi(-\tfrac{z}{2})\gamma^\mu\mathcal W(-\tfrac{z}{2},\tfrac{z}{2}|n)\psi(\tfrac{z}{2})|P,S\rangle\Big|_{z^+=0}.
\end{equation}
Its parametrization in terms of canonical twist-2, twist-3 and twist-4 quark TMDs reads~\cite{Goeke:2005hb,Bacchetta:2006tn}
\begin{equation}
\begin{aligned}
\Phi^{[ \gamma^{+} ]}&=f_1- \frac{\epsilon_T^{k_T  S_T}}{M}\, f^{\perp}_{1T},  \\
\Phi^{[ \gamma^{\alpha}_T ]}&= \frac{M}{P^+}
\left[ \frac{k_T^\alpha}{M}\,f^{\perp} - \epsilon_T^{\alpha S_T}  f_T
-  \lambda\, \frac{\epsilon_T^{\alpha k_T}}{M}\,
 f^{\perp}_{L}
- \frac{k_T^\alpha k_T^\beta-\frac{1}{2} k_T^2 g_T^{\alpha\beta} }{M^2}
\,\epsilon_{T\beta S_T} f^{\perp}_{T}  \right],  \\
\Phi^{[ \gamma^{-} ]}&=\left(\frac{M}{P^+}\right)^{\!2}
\left[f_3-  \frac{\epsilon_T^{k_T  S_T}}{M}\, f^{\perp}_{3T} \right],
\label{eqn:tmd1}
\end{aligned}
\end{equation}
reminding that $k^2_T=-\uvec k_\perp^2$.  TMDs are scale-dependent objects\footnote{Beyond canonical twist-2, the renormalization of TMDs is troublesome and the evolution equations are not closed, see Ref.~\cite{Rodini:2022wki}.} that are extracted from fits to experimental data. Their QCD evolution has been a major focus of the past decade and is expected to play a significant role at the future Electron-Ion Collider~\cite{Angeles-Martinez:2015sea}.  Note that the twist-2 functions $f_1$ and $f_{1T}^\perp$ are often referred to as the ``unpolarized'' TMDs in the literature.

Setting $\nu=+$ in Eq.~\eqref{reltoTMD}, we find
\begin{equation}
    \begin{aligned}
a_1&=xf_1,  &a_{1T}^\perp&=xf^{\perp}_{1T}, \\
\tfrac{1}{2}a_1+a_5&=xf_3,  &\qquad \tfrac{1}{2}a_{1T}^\perp+a_{5T}^\perp&=xf_{3T}^\perp,\\
a_7&=x f^{\perp}, & a_{7T}^\perp&=xf^{\perp}_{T}, \\
a_{2L}&=x f^{\perp}_{L} , & a_{2T}&=x f^+_{T},
\label{eqn:rela-lf}
\end{aligned}
\end{equation}
where $f^\pm_{T}=f_{T} \pm\tfrac{\uvec k_\perp^2}{2M^2}\,f^{\perp}_{T}$. Similarly, setting $\nu=i\in\{1,2\}$ in Eq.~\eqref{reltoTMD} gives
\begin{equation}
    \begin{aligned}
a_3&=f^\perp,  &a_{3T}^\perp&=f^{\perp}_T, \\
a_6&=f_1, & a_{6T}^\perp&=f^{\perp}_{1T}, \\
\tfrac{1}{2}a_6+a_8&=f_{3},  &\tfrac{1}{2}a_{6T}^\perp+a_{8T}^\perp&=f^{\perp}_{3T},\\
a_{6L}&=f^\perp_{L} , & a_{6T}&=f^+_T,\\
a_0&=a_{1L}=a_{3L}=a_{5L}=0,& \qquad a_{0T}^\perp&=a_{1T}=a_{3T}=a_{5T}=0.
\label{eqn:rela-gic}
\end{aligned}
\end{equation}
These relations imply that the quark TMD EMT involves only 16 independent functions
\begin{equation}
\begin{aligned}
\mathcal{T}^{\mu \nu}_q= \frac{1}{P^+} \Bigg\{& \left[\tilde P^{\mu}f_1+k_T^{\mu}f^\perp+N^{\mu}f_3-\frac{\epsilon_{T}^{k_T  S_T } }{M} \left( \tilde P^{\mu}f_{1T}^\perp+k_T^{\mu}f_{T}^\perp
+N^{\mu}f^\perp_{3T}\right)-M\epsilon_{T}^{\mu  S_T } f^+_{T}-\lambda\epsilon_{T}^{\mu  k_T } f_{L}^\perp\right]\tilde k^{\nu}\\
&+\left[\tilde P^{\mu}\check f_1+k_T^{\mu}\check f^\perp+N^{\mu}\check f_3-\frac{\epsilon_{T}^{k_T  S_T } }{M} \left(\tilde P^{\mu}\check f_{1T}^\perp+k_T^{\mu}\check f_{T}^\perp+N^{\mu}\check f_{3T}^\perp\right)-M\epsilon_{T}^{\mu  S_T } \check f^+_{T}-\lambda\epsilon_{T}^{\mu  k_T } \check f_{L}^\perp\right]N^{\nu}\Bigg\},
\label{eqn:emtgic}
\end{aligned}
\end{equation}
where we introduce for convenience $\tilde P^\mu=[P^+,0,\uvec 0_\perp]$ and $\tilde k^\mu=[xP^+,0,\uvec k_\perp]$. The combinations
\begin{equation}
\begin{aligned}
    \check f_1&=a_4+\tfrac{x}{2}f_1,&\check f_{1T}^\perp&=a_{4T}^\perp+\tfrac{x}{2}f_{1T}^\perp,\\
    \check f^\perp&=a_9+\tfrac{x}{2}f^\perp,&\tilde f_{T}^\perp&=a_{9T}^\perp+\tfrac{x}{2}f_{T}^\perp,\\
    \check f_3&=a_2+\tfrac{1}{2}a_4+\tfrac{x}{2}f_3,&\qquad \check f^\perp_{3T}&=a_{2T}^\perp+\tfrac{1}{2}a_{4T}^\perp+\tfrac{x}{2}f^\perp_{3T},\\
    \check f^\perp_L&=a_{4L}+\tfrac{x}{2}f_L^\perp,&\check f^+_{T}&=a_{4T}+\tfrac{x}{2}f^+_{T},
\end{aligned}
\end{equation}
parametrize the information that cannot be accessed with the ordinary quark vector TMDs.

\section{Mechanical properties}
\label{sect4}

The interpretation of the light-front components of the EMT and the associated distributions in impact-parameter space have been discussed in Refs.~\cite{Lorce:2018egm,Freese:2021czn}. We investigate here their momentum-space counterparts.

\subsection{Densities of longitudinal and transverse momentum}

Since the TMD correlator $\Phi^{[\gamma^+]}(P,x,\uvec k_\perp,N,S;\eta)$ is interpreted as the probability density of finding a quark with three-momentum $[xP^+,\uvec k_\perp]$, it is natural to interpret
\begin{equation}\label{momdens}
\begin{aligned}
    \mathcal T^{++}_q&=xP^+\Phi^{[\gamma^+]}=\left(f_1- \frac{\epsilon_T^{k_T  S_T}}{M}\, f^{\perp}_{1T}\right)xP^+,\\
    \mathcal T^{+i}_q&=k^i_T\Phi^{[\gamma^+]}=\left(f_1- \frac{\epsilon_T^{k_T  S_T}}{M}\, f^{\perp}_{1T}\right)k^i_T,\qquad\quad i=1,2
\end{aligned}
\end{equation}
as the quark longitudinal and transverse momentum densities in momentum space. The average quark longitudinal momentum is then obtained by integration over the quark momentum
\begin{equation}
    \langle k^+\rangle_q=\langle x\rangle_qP^+=\int\ud x\,\ud^2k_\perp\,\mathcal T^{++}_q=P^+\int\ud x\,\ud^2k_\perp\,xf_1.
\end{equation}
Similarly, the average quark transverse momentum~\cite{Boer:2003cm,Burkardt:2003yg,Meissner:2007rx,Amor-Quiroz:2020qmw} is given by
\begin{equation}
    \langle k^i_\perp\rangle_q=\int\ud x\,\ud^2k_\perp\,\mathcal T^{+i}_q=\epsilon_T^{iS_T}\int\ud x\,\ud^2k_\perp\,\frac{\uvec k^2_\perp}{2M}\,f_{1T}^\perp.
\end{equation}

\begin{figure}[h]
	\includegraphics[angle=0,scale=0.5]{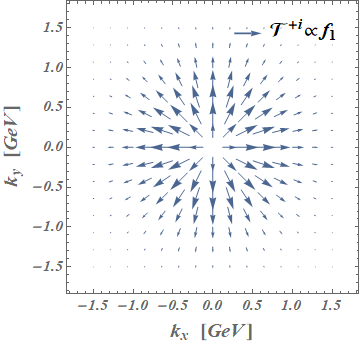}\hspace{0.75cm}
 \includegraphics[angle=0,scale=0.5]{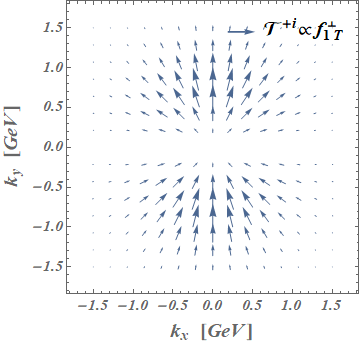}
	\caption{Illustration of the contributions to the quark transverse momentum density inside a nucleon polarized along the $x$-axis, using a simple gaussian model for the transverse momentum dependence.}
	\label{Fig01}
\end{figure}

We illustrate in Fig.~\ref{Fig01} the two contributions to the transverse momentum density at some fixed value of $x$ using a simple gaussian model for the transverse momentum dependence $f(\uvec k^2_\perp)\propto e^{-\uvec k_\perp^2/\langle\uvec k^2_\perp\rangle}$ with the typical value $\langle\uvec k^2_\perp\rangle\approx 0.6$ GeV$^2$ for the gaussian width~\cite{Anselmino:2013lza}. Since $\mathcal T^{+i}_q\propto k^i_T$, it is natural that the transverse momentum density looks like a hedgehog. The unpolarized contribution driven by $f_1$ is necessarily axially symmetric for there is no preferred transverse direction. However, the combination of target transverse polarization and initial/final state interactions breaks axial symmetry. The magnitude of this effect is quantified by the Sivers function $f^\perp_{1T}$~\cite{Sivers:1989cc}.

\subsection{Transverse flux of inertia}

Because of the Galilean subgroup exhibited by the light-front coordinates, the light-front longitudinal momentum plays the role of inertia in the transverse plane~\cite{Susskind:1967rg}. The transverse flux of longitudinal momentum $\mathcal T^{i+}$ can therefore be thought of as the transverse flux of inertia, suggesting the definition of an effective quark transverse velocity via the ratio
\begin{equation}
    v^i_\perp=\frac{\mathcal T^{i+}_q}{\mathcal T^{++}_q}.
\end{equation}
It is often thought that the EMT is symmetric, and hence that momentum density $\mathcal T^{+i}_q$ equals flux of inertia $\mathcal T^{i+}_q$. In that case, the quark transverse velocity is simply given by $\uvec v_\perp=\uvec k_\perp/(xP^+)$. In a gauge theory, velocity and canonical momentum are however usually not parallel and we should expect in general\footnote{Quark spin may also make the EMT asymmetric, but the antisymmetric contribution vanishes when initial and final target momenta are the same~\cite{Leader:2013jra}.} $\mathcal T^{i+}_q\neq \mathcal T^{+i}_q$. Indeed, we find that
\begin{equation}\label{inertiaflux}
   \mathcal T^{i+}_q=xP^+\Phi^{[\gamma^i_T]}=\left(xf^\perp- \frac{\epsilon_T^{k_T  S_T}}{M}\, xf^{\perp}_{T}\right)k^i_T- M\epsilon_T^{i S_T}  xf^+_T-\lambda \epsilon_T^{i k_T}
 xf^{\perp}_{L}.
\end{equation}
In the case of a symmetric TMD EMT, we should have
\begin{equation}
    \begin{aligned}
        xf^\perp&=f_1,\\
        xf^\perp_T&=f^\perp_{1T},\\
        f^+_T&=f^\perp_L=0.
    \end{aligned}
\end{equation}
Interestingly, the first relation was found in Ref.~\cite{Lorce:2014hxa} using the free quark equation of motion. In QCD, these relations are not expected to hold in general and their violations are a direct measure of the interaction between quarks and gluons.

\begin{figure}[h]
	\includegraphics[angle=0,scale=0.5]{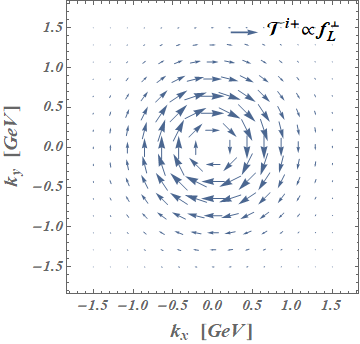}\hspace{0.75cm}
 \includegraphics[angle=0,scale=0.5]{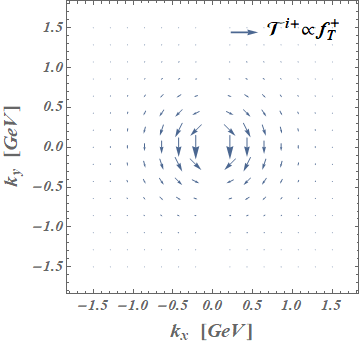}
	\caption{Illustration of two contributions to the quark transverse flux of inertia inside a nucleon polarized along the $z$-axis (left panel) or the $x$-axis (right panel), using a simple gaussian model for the transverse momentum dependence. The other two contributions are similar to those given in Fig.~\ref{Fig01}.}
	\label{Fig02}
\end{figure}

Decomposing $\epsilon_T^{i S_T}$ onto components parallel and orthogonal to $k^i_\perp$, we can rewrite Eq.~\eqref{inertiaflux} as
\begin{equation}
   \mathcal T^{i+}_q=\left(xf^\perp- \frac{M\epsilon_T^{k_T  S_T}}{k^2_T}\, xf^-_T\right)k^i_T-\left(
 \lambda\, xf^{\perp}_{L}+\frac{M(k_T\cdot S_T)}{k^2_T}\,xf^+_T\right)\epsilon_T^{i k_T}.
\end{equation}
The first two terms have the same structure as $\mathcal T^{+i}_q$ in Eq.~\eqref{momdens} and lead to similar hedgehog distributions as in Fig.~\ref{Fig01}. The last two terms indicate that besides modifying the magnitude of the quark velocity, QCD interactions can also modify its direction relative to $\uvec k_\perp$. The corresponding distributions are illustrated in Fig.~\ref{Fig02} using the same simple gaussian model as for $\mathcal T^{+i}_q$.

\subsection{Transverse pressure and shear forces}

The notions of 2D spatial distributions of pressure (or isotropic stress) $\sigma$ and shear forces (or pressure anisotropy) $\Pi$ have been introduced in Ref.~\cite{Lorce:2018egm}. Similarly, we introduce here the distributions of transverse pressure and shear forces in momentum space
\begin{equation}\label{2Dstress}
   \mathcal T^{ij}=-g^{ij}_T\,\sigma+\left(\frac{1}{2}\,g^{ij}_T-\frac{k^i_T k^j_T}{k^2_T}\right)\Pi+\frac{k^i_T\epsilon^{jk_T}+k^j_T\epsilon^{ik_T}}{2 k^2_T}\,\Pi^S+\epsilon^{ij}_T\,\Pi^A.
\end{equation}
The first two transverse tensors are similar to those found in position space, with transverse momentum $\uvec k_\perp$ replacing impact parameter $\uvec b_\perp$. The last two transverse tensors are new, and are allowed provided that $\Pi_S$ and $\Pi_A$ are linear in the target polarization and naive \textsf{T}-odd. Using the particular structure of the quark EMT
\begin{equation}\label{2Dstress-quark}
\begin{aligned}
   \mathcal T^{ij}_q=k^j_T\Phi^{[\gamma^i_T]}&=\tfrac{k^j_T}{xP^+}\,\mathcal T^{i+}_q\\
   &=\frac{1}{P^+}\left[\left(f^\perp- \frac{M\epsilon_T^{k_T  S_T}}{k^2_T}\, f^-_T\right)k^i_T k^j_T-\left(
 \lambda\, f^{\perp}_{L}+\frac{M(k_T\cdot S_T)}{k^2_T}\,f^+_T\right)\epsilon_T^{i k_T}k^j_T\right],
 \end{aligned}
\end{equation}
we find
\begin{equation}
    \begin{aligned}
        \sigma_q&=\tfrac{1}{2}\Pi_q=-\frac{1}{2P^+}\left[k^2_T f^\perp-M\epsilon^{k_TS_T}f^-_T\right],\\
        \Pi^A_q&=\tfrac{1}{2}\Pi^S_q=-\frac{1}{2P^+}\left[ \lambda\, k^2_T f^{\perp}_{L}+M( k_T\cdot S_T)f^+_T\right].
    \end{aligned}
\end{equation}
This is to be compared with the free quark case given by $\mathcal T^{ij}_{q,\text{free}}=f_1\, k^i_T k^j_T/(xP^+)$.

\section{Summary}
\label{sect5}

The energy-momentum tensor is a fundamental object in any relativistic field theory. In hadronic physics, it provides key information about quark and gluon contributions to the nucleon mass and spin, and is therefore at the heart of the physics program of the forthcoming Electron-Ion Collider in the US.

In this work we introduced the concept of energy-momentum tensor distribution in momentum space. In the case of a spin-$0$ target, we found that this distribution can be parametrized in terms of 10 independent gravitational transverse-momentum distributions. For a spin-$\frac{1}{2}$ target, we obtained in general 32 independent functions. Due to the particular structure of the gauge-invariant canonical energy-momentum tensor for quarks, this number reduces to 16, half of which can directly be expressed in terms of the usual quark vector transverse-momentum distributions. A similar analysis can in principle be applied to the gluon energy-momentum tensor, but is left for a future dedicated investigation. We discussed the physical interpretation of various components of the energy-momentum tensor and we used a simple gaussian model for illustration. We observed in particular that the stress tensor distribution in momentum space is expected to be asymmetric due to spin-dependent contributions associated with initial/final-state interactions.

At the present stage, only a few gravitational transverse-momentum distributions can be extracted from actual experiments. Our work provides however new motivations for studying and measuring higher-twist transverse-momentum distributions. In the meantime, it will be interesting to investigate these gravitational transverse-momentum distributions within other approaches, such as Lattice QCD and model calculations.

\section*{Acknowledgements}
We thank Simone Rodini for drawing our attention to recent developments regarding the status of higher-twist transverse-momentum distributions. Qin-Tao Song was supported by the National Natural Science Foundation of China under Grant Number 12005191 and the China Scholarship Council
for visiting \'Ecole polytechnique.


\begin{thebibliography}{99}

\bibitem{Burkert:2023wzr}
V.~D.~Burkert, L.~Elouadrhiri, F.~X.~Girod, C.~Lorc\'e, P.~Schweitzer and P.~E.~Shanahan,
[arXiv:2303.08347 [hep-ph]].

\bibitem{Ji:1994av}
X.~D.~Ji,
Phys. Rev. Lett. \textbf{74}, 1071-1074 (1995).

\bibitem{Ji:1995sv}
X.~D.~Ji,
Phys. Rev. D \textbf{52}, 271-281 (1995).

\bibitem{Yang:2018nqn}
Y.~B.~Yang, J.~Liang, Y.~J.~Bi, Y.~Chen, T.~Draper, K.~F.~Liu and Z.~Liu,
Phys. Rev. Lett. \textbf{121}, no.21, 212001 (2018).

\bibitem{Hatta:2018sqd}
Y.~Hatta, A.~Rajan and K.~Tanaka,
JHEP \textbf{12}, 008 (2018).

\bibitem{Lorce:2017xzd}
C.~Lorc\'e,
Eur. Phys. J. C \textbf{78}, no.2, 120 (2018).

\bibitem{Metz:2020vxd}
A.~Metz, B.~Pasquini and S.~Rodini,
Phys. Rev. D \textbf{102}, 114042 (2020).

\bibitem{Lorce:2021xku}
C.~Lorc\'e, A.~Metz, B.~Pasquini and S.~Rodini,
JHEP \textbf{11}, 121 (2021).

\bibitem{Jaffe:1989jz}
R.~L.~Jaffe and A.~Manohar,
Nucl. Phys. B \textbf{337}, 509-546 (1990).

\bibitem{Ji:1996ek}
X.~D.~Ji,
Phys. Rev. Lett. \textbf{78}, 610-613 (1997).

\bibitem{Leader:2013jra}
E.~Leader and C.~Lorc\'e,
Phys. Rept. \textbf{541}, no.3, 163-248 (2014).

\bibitem{Wakamatsu:2014zza}
M.~Wakamatsu,
Int. J. Mod. Phys. A \textbf{29}, 1430012 (2014).

\bibitem{Lorce:2021gxs}
C.~Lorc\'e,
Eur. Phys. J. C \textbf{81}, no.5, 413 (2021).

\bibitem{Polyakov:2002yz}
M.~V.~Polyakov,
Phys. Lett. B \textbf{555}, 57-62 (2003).

\bibitem{Polyakov:2018zvc}
M.~V.~Polyakov and P.~Schweitzer,
Int. J. Mod. Phys. A \textbf{33}, no.26, 1830025 (2018).

\bibitem{Burkert:2018bqq}
V.~D.~Burkert, L.~Elouadrhiri and F.~X.~Girod,
Nature \textbf{557}, no.7705, 396-399 (2018).

\bibitem{Lorce:2018egm}
C.~Lorc\'e, H.~Moutarde and A.~P.~Trawi\'nski,
Eur. Phys. J. C \textbf{79}, no.1, 89 (2019).

\bibitem{Freese:2021czn}
A.~Freese and G.~A.~Miller,
Phys. Rev. D \textbf{103}, 094023 (2021).

\bibitem{AbdulKhalek:2021gbh}
R.~Abdul Khalek, A.~Accardi, J.~Adam, D.~Adamiak, W.~Akers, M.~Albaladejo, A.~Al-bataineh, M.~G.~Alexeev, F.~Ameli and P.~Antonioli, \textit{et al.}
Nucl. Phys. A \textbf{1026}, 122447 (2022).

\bibitem{AbdulKhalek:2022hcn}
R.~Abdul Khalek, U.~D'Alesio, M.~Arratia, A.~Bacchetta, M.~Battaglieri, M.~Begel, M.~Boglione, R.~Boughezal, R.~Boussarie and G.~Bozzi, \textit{et al.}
[arXiv:2203.13199 [hep-ph]].

\bibitem{Kobzarev:1962wt}
I.~Y.~Kobzarev and L.~B.~Okun,
Zh. Eksp. Teor. Fiz. \textbf{43}, 1904-1909 (1962).

\bibitem{Pagels:1966zza}
H.~Pagels,
Phys. Rev. \textbf{144}, 1250-1260 (1966).

\bibitem{Bakker:2004ib}
B.~L.~G.~Bakker, E.~Leader and T.~L.~Trueman,
Phys. Rev. D \textbf{70}, 114001 (2004).

\bibitem{Cotogno:2019vjb}
S.~Cotogno, C.~Lorc\'e, P.~Lowdon and M.~Morales,
Phys. Rev. D \textbf{101}, no.5, 056016 (2020).

\bibitem{Kumano:2017lhr}
S.~Kumano, Q.~T.~Song and O.~V.~Teryaev,
Phys. Rev. D \textbf{97}, no.1, 014020 (2018).

\bibitem{Diehl:2003ny}
M.~Diehl,
Phys. Rept. \textbf{388}, 41-277 (2003).

\bibitem{Lorce:2015lna}
C.~Lorc\'e,
JHEP \textbf{08}, 045 (2015).

\bibitem{Guo:2021aik}
Y.~Guo, X.~Ji and K.~Shiells,
Nucl. Phys. B \textbf{969}, 115440 (2021).

\bibitem{Boer:1997nt}
D.~Boer and P.~J.~Mulders,
Phys. Rev. D \textbf{57}, 5780-5786 (1998).

\bibitem{Burkardt:2003yg}
M.~Burkardt,
Phys. Rev. D \textbf{69}, 057501 (2004).

\bibitem{Burkardt:2004ur}
M.~Burkardt,
Phys. Rev. D \textbf{69}, 091501 (2004).

\bibitem{Boer:2015vso}
D.~Boer, C.~Lorc\'e, C.~Pisano and J.~Zhou,
Adv. High Energy Phys. \textbf{2015}, 371396 (2015).

\bibitem{Amor-Quiroz:2020qmw}
D.~A.~Amor-Quiroz, M.~Burkardt, W.~Focillon and C.~Lorc\'e,
Eur. Phys. J. C \textbf{81}, no.7, 589 (2021).

\bibitem{Meissner:2009ww}
S.~Meissner, A.~Metz and M.~Schlegel,
JHEP \textbf{08}, 056 (2009).

\bibitem{Bacchetta:2006tn}
A.~Bacchetta, M.~Diehl, K.~Goeke, A.~Metz, P.~J.~Mulders and M.~Schlegel,
JHEP \textbf{02}, 093 (2007).

\bibitem{Lorce:2012ce}
C.~Lorc\'e,
Phys. Lett. B \textbf{719}, 185-190 (2013).

\bibitem{Belitsky:2002sm}
A.~V.~Belitsky, X.~Ji and F.~Yuan,
Nucl. Phys. B \textbf{656}, 165-198 (2003).

\bibitem{Lorce:2012rr}
C.~Lorc\'e,
Phys. Rev. D \textbf{87}, no.3, 034031 (2013).

\bibitem{Chen:2008ag}
X.~S.~Chen, X.~F.~Lu, W.~M.~Sun, F.~Wang and T.~Goldman,
Phys. Rev. Lett. \textbf{100}, 232002 (2008).

\bibitem{Wakamatsu:2010cb}
M.~Wakamatsu,
Phys. Rev. D \textbf{83}, 014012 (2011).

\bibitem{Hatta:2011zs}
Y.~Hatta,
Phys. Rev. D \textbf{84}, 041701 (2011).

\bibitem{Hatta:2011ku}
Y.~Hatta,
Phys. Lett. B \textbf{708}, 186-190 (2012).

\bibitem{Ji:2003ak}
X.~d.~Ji,
Phys. Rev. Lett. \textbf{91}, 062001 (2003).

\bibitem{Belitsky:2003nz}
A.~V.~Belitsky, X.~d.~Ji and F.~Yuan,
Phys. Rev. D \textbf{69}, 074014 (2004).

\bibitem{Lorce:2011kd}
C.~Lorc\'e and B.~Pasquini,
Phys. Rev. D \textbf{84}, 014015 (2011).

\bibitem{Lorce:2011ni}
C.~Lorc\'e, B.~Pasquini, X.~Xiong and F.~Yuan,
Phys. Rev. D \textbf{85}, 114006 (2012).

\bibitem{Goeke:2005hb}
K.~Goeke, A.~Metz and M.~Schlegel,
Phys. Lett. B \textbf{618}, 90-96 (2005).

\bibitem{Rodini:2022wki}
S.~Rodini and A.~Vladimirov,
JHEP \textbf{08}, 031 (2022)
[erratum: JHEP \textbf{12}, 048 (2022)].

\bibitem{Angeles-Martinez:2015sea}
R.~Angeles-Martinez, A.~Bacchetta, I.~I.~Balitsky, D.~Boer, M.~Boglione, R.~Boussarie, F.~A.~Ceccopieri, I.~O.~Cherednikov, P.~Connor and M.~G.~Echevarria, \textit{et al.}
Acta Phys. Polon. B \textbf{46}, no.12, 2501-2534 (2015).

\bibitem{Boer:2003cm}
D.~Boer, P.~J.~Mulders and F.~Pijlman,
Nucl. Phys. B \textbf{667}, 201-241 (2003).

\bibitem{Meissner:2007rx}
S.~Meissner, A.~Metz and K.~Goeke,
Phys. Rev. D \textbf{76}, 034002 (2007).

\bibitem{Anselmino:2013lza}
M.~Anselmino, M.~Boglione, J.~O.~Gonzalez Hernandez, S.~Melis and A.~Prokudin,
JHEP \textbf{04}, 005 (2014).

\bibitem{Sivers:1989cc}
D.~W.~Sivers,
Phys. Rev. D \textbf{41}, 83 (1990).

\bibitem{Susskind:1967rg}
L.~Susskind,
Phys. Rev. \textbf{165}, 1535-1546 (1968).

\bibitem{Lorce:2014hxa}
C.~Lorc\'e, B.~Pasquini and P.~Schweitzer,
JHEP \textbf{01}, 103 (2015).



\end{thebibliography}
\end{document}